\documentclass[twocolumn,showpacs,preprintnumbers,amsmath,amssymb]{revtex4}

\usepackage{graphicx}
\usepackage{dcolumn}
\usepackage{bm}

\usepackage{color} 

\begin{document}

\title{Crossover between distinct mechanisms of microwave photoresistance in bilayer systems}

\author{ S. Wiedmann,$^{1,2}$ G. M. Gusev,$^3$ O. E. Raichev,$^4$ A. K. Bakarov,$^5$
and J. C. Portal$^{1,2,6}$} 
 \affiliation{$^1$LNCMI-CNRS, UPR 3228, BP
166, 38042 Grenoble Cedex 9, France} \affiliation{$^2$INSA Toulouse,
31077 Toulouse Cedex 4, France} \affiliation{$^3$Instituto de
F\'{\i}sica da Universidade de S\~ao Paulo, CP 66318 CEP 05315-970,
S\~ao Paulo, SP, Brazil}\affiliation{$^4$Institute of
Semiconductor Physics, NAS of Ukraine, Prospekt Nauki 41, 03028,
Kiev, Ukraine} \affiliation{$^5$Institute of Semiconductor Physics,
Novosibirsk 630090, Russia}\affiliation{$^6$Institut Universitaire
de France, 75005 Paris, France}

\date{\today}

\begin{abstract}
We report on temperature-dependent magnetoresistance measurements in balanced double 
quantum wells exposed to microwave irradiation for various frequencies. We have found 
that the resistance oscillations are described by the microwave-induced modification of 
electron distribution function limited by inelastic scattering (inelastic mechanism), 
up to a temperature of $T^{*} \simeq 4$~K. With increasing temperature, a strong 
deviation of the oscillation amplitudes from the behavior predicted by this mechanism 
is observed, presumably indicating a crossover to another mechanism of microwave 
photoresistance, with similar frequency dependence. Our analysis shows that this 
deviation cannot be fully understood in terms of contribution from the mechanisms 
discussed in theory. 
\end{abstract}


\maketitle

\section{Introduction}

The physics of two-dimensional (2D) electron systems exposed to a continuous microwave 
irradiation in the presence of perpendicular magnetic fields $B$ has attracted both experimental and 
theoretical attention in the last years following the observation of the microwave-induced resistance 
oscillations (MIROs) \cite{1} which evolve into ``zero resistance states" (ZRS) \cite{2,3} for a 
sufficiently high microwave intensity. The MIRO periodicity is governed by the ratio of the radiation 
frequency $\omega$ to the cyclotron frequency $\omega_{c}=eB/m$, where $m$ is the effective 
mass of the electrons. These oscillations occur because of Landau quantization and originate 
from the scattering-assisted electron transitions between different Landau levels, which 
become possible in the presence of microwave excitation. Two competing microscopic mechanisms 
of the oscillating photoresistance have been proposed theoretically: the ``displacement" 
mechanism which accounts for spatial displacement of electrons along the applied dc field 
under scattering-assisted microwave absorption \cite{4,5}, and ``inelastic" mechanism, 
owing to an oscillatory contribution to the isotropic part of the electron distribution function 
\cite{6,7}. Both mechanisms describe phase and periodicity of MIROs observed 
in experiments. A systematic theoretical study of photoresistance has revealed two 
additional mechanisms: the ``quadrupole" mechanism, which comes from excitation of the 
second angular harmonic of the distribution function, and ``photovoltaic" mechanism, 
which is described as a combined action of the microwave and dc fields on both temporal 
and angular harmonics of the distribution function \cite{7}. Both additional mechanisms 
contribute to transverse (Hall) dc resistance, while the photovoltaic mechanism contributes 
also to diagonal resistance. However, this contribution is found to be weak and has not 
been detected in MIROs observed in experiments.

For low temperatures the inelastic mechanism plays the dominant role 
because the relaxation of the microwave-induced oscillatory part of the electron distribution 
is slow. This relaxation is governed by the inelastic electron-electron scattering with 
a characteristic time $\tau_{in} \propto T^{-2}$, which is in the order of 1 ns at 
temperatures $T \simeq 1$~K. This $T^{-2}$-dependence has also been found
experimentally in Ref. 8. 
Nevertheless, recent experiments on high-mobility samples 
suggest that the displacement mechanism cannot be ignored and becomes important with 
increasing temperature, when the relative contribution of the inelastic mechanism decreases 
\cite{9}. The crossover between these two mechanisms was observed at $T \simeq 2$ K.
Notice that, since these mechanisms produce nearly the same frequency dependence of 
MIROs, the only way to distinguish between them is to measure temperature dependence 
of the oscillation amplitudes. For a better understanding of the role of inelastic and 
displacement mechanisms in microwave-induced resistance of 2D electrons, systematic 
experiments in different samples are highly desirable. 
  
In this paper we undertake a study of temperature dependence of magnetoresistance in 
two-subband electron systems formed in double quantum wells (DQWs). Recently, we have 
found \cite{10} that the inelastic mechanism satisfactorily explains low-temperature 
photoresistance in such systems exposed to microwave irradiation. The main difference 
in magnetoresistance of two-subband electron systems with respect to conventional 
(single-subband) 2D systems is the presence of magneto-intersubband (MIS) oscillations 
(see Refs. \cite{11},\cite{12},\cite{13} and references therein) which occur owing to periodic modulation 
of the probability of intersubband transitions by the magnetic field. Under microwave 
irradiation, these oscillations interfere with MIROs. The interference causes a peculiar 
magnetoresistance picture where one may see enhancement, suppression, or inversion 
(flip) of MIS peaks, correlated with the microwave frequency \cite{10}. Whereas such a behavior 
of magnetoresistance is more complicated than that for single-subband electron systems, 
it offers certain advantages in analyzing the effect of microwaves. The reason is that 
the quantum component of magnetoresistance, which is affected by the microwaves, 
is ``visualized" in DQWs by the MIS oscillations whose period is typically smaller than 
the period of the MIROs. As a result, the changes in MIRO amplitudes caused by variation 
in temperature or microwave intensity can be traced by observation of the behavior of 
single MIS peaks, and the position of node points of the MIROs can be determined more 
distinctly by the MIS peak inversion.   

Our main result can be summarized as follows. We find that the inelastic mechanism 
fails to explain the observed photoresistance for $T>4$~K. The temperature dependence 
of magnetoresistance can be explained either by a deviation from the $\tau_{in} \propto T^{-2}$
law at these temperatures or by inclusion of another, $T$-independent contribution 
to MIROs. The first possibility seems to be unlikely, because we see no reasons for 
such a deviation. The second possibility is more promising, and a consideration of an 
additional contribution owing to the displacement mechanism seems to be a natural choice. 
However, our quantitative estimates demonstrate that the crossover from the inelastic 
to the displacement mechanism of MIROs is expected at higher temperatures in our samples, 
around 10~K. Therefore, the origin of the observed photoresistance behavior 
can be partially explained by a contribution of displacement mechanism but does not fully
account for our finding.

The paper is organized as follows. In Sec. II we present details of the experimental
analysis and the theoretical consideration of the microwave-induced resistivity of 
two-subband systems. In Sec. III we analyze the deviation from the inelastic mechanism 
with increasing temperature, compare our experimental results with the theory 
including both inelastic and displacement mechanisms, and formulate our conclusions. 

\section{Experimental and theoretical basis}

We have studied balanced GaAs DQWs separated by different Al$_{x}$Ga$_{1-x}$As barriers
with barrier thicknesses of $d_{b}$=14, 20 and 30 \AA \ in perpendicular magnetic fields. 
We have analyzed two wafers with $d_{b}$=14 \AA \ and we
focus in this paper on the samples with subband separation of $\Delta$~=~3.05~meV. This value is 
extracted from the periodicity of low-field MIS oscillations. The samples have a high total sheet 
electron density $n_s \simeq 1.15 \times 10^{12}$ cm$^{-2}$ and a mobility of $\mu~\simeq 1.4 
\times 10^{6}$ cm$^{2}$/V s at 1.4~K. The measurements have been carried out in a
VTI cryostat using conventional lock-in technique to measure the longitudinal resistance $R~=~R_{xx}$ 
under a continuous microwave irradiation. As MW sources, we employ different ``carcinotron" generators 
and we focus on the frequency range between 55 and 140~GHz. A circular-section waveguide delivers 
microwave radiation down to the sample which is placed at a distance of 1-2~mm in front of the waveguide 
output.

\begin{figure}[ht]
\includegraphics[width=9cm]{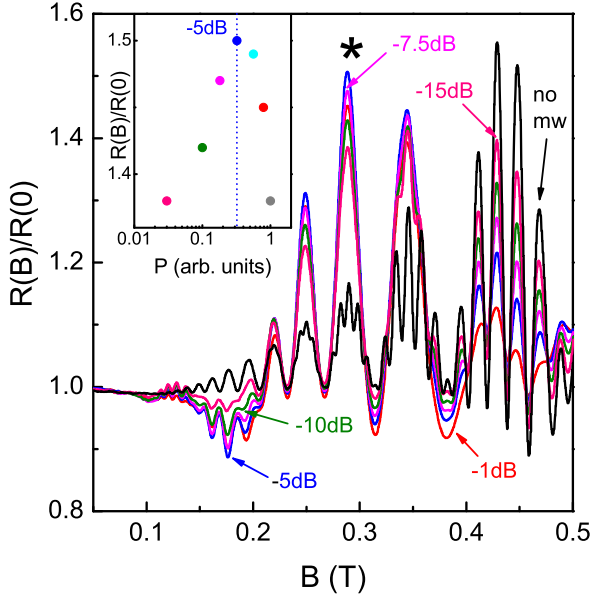}
\caption{\label{fig1} (Color online) Normalized power dependent photoresistance as a 
function of the magnetic field for 85~GHz at $T$~=~1.4~K. Without microwave irradiation (no mw), 
MIS oscillations are visible, superimposed on SdH oscillations. An increase in microwave intensity 
leads to an enhancement, damping, or flip of MIS peaks. We observe a saturation of the 
MIS oscillation for the attenuations between -2.5 and -5~dB; the inset shows the 
amplitude of the MIS peak marked with an asterisk.}
\end{figure}

In Fig. \ref{fig1} we present the basis of our experimental analysis for further temperature 
dependent measurements. Without microwaves (no mw), we observe MIS oscillations which are 
superimposed on low-field Shubnikov-de Haas (SdH) oscillations at low temperatures. As the microwave 
power increases (at a fixed microwave frequency of 85~GHz), the MIS oscillation picture is 
modified by the MIRO contribution. It is worth mentioning that we have to perform the experimental 
analysis for low microwave intensity to ensure that the amplitude of MIS peaks is not yet saturated. 
Thus we present in Fig. \ref{fig1} power dependent measurements for several chosen attenuations: 
0, -1, -2.5, -5, -7.5, -10 and -15~dB. The inset to Fig. \ref{fig1} shows MIS peak amplitude at 
$B$~=~0.3~T (marked by an asterisk) where saturation occurs between -2.5 and -5~dB. Therefore, we 
use experimental data with lower microwave intensity (for this frequency $P~\leq$~-7.5~dB). Still, 
the heating of 2D electrons by microwaves is observable at these intensities by a suppression 
of SdH oscillations. This heating is not strong and does not lead to the bolometric effect 
at $\omega_{c} \simeq \omega$ because of the radiative broadening of the cyclotron resonance 
\cite{8}, \cite{14}. For temperatures below 10 K the phonon-induced contribution to electron 
mobility in our samples is weak, so the transport is controlled by the electron-impurity 
scattering. 

Our theoretical model takes into account both inelastic and displacement mechanisms of 
photoresistance generalized to the two-subband case (for generalization to an arbitrary 
number of subbands, see Ref. 15). In the regime of classically strong magnetic fields, 
the symmetric part of the diagonal resistivity, $\rho_{d}$, in the presence of microwaves 
is given by the expression
\begin{eqnarray}
\frac{\rho_{d}}{\rho_{0}}= 1 - 2 {\cal T} \tau_{tr} \sum_{j=1,2} \nu^{tr}_j 
d_j \cos \frac{2 \pi (\varepsilon_F -\varepsilon_j)}{\hbar \omega_c} \nonumber \\ 
+\tau_{tr} \biggl[ \sum_{j=1,2} \frac{2n_j}{n_s} \nu^{tr}_{jj} d_j^2 +  
2 \nu^{tr}_{12} d_1 d_2 \cos \frac{2 \pi \Delta}{\hbar \omega_c} \biggr] \nonumber \\
-\frac{1}{2} \tau^2_{tr} A_{\omega} \biggl[ \sum_{j=1,2} (\nu^{tr}_{j} d_j)^2 +  
2 \nu^{tr}_{1} \nu^{tr}_{2} d_1 d_2 \cos \frac{2 \pi \Delta}{\hbar \omega_c} \biggr] \nonumber \\
-\tau^* B_{\omega} \biggl[ \sum_{j=1,2} \left(\frac{2n_j}{n_s} \right)^2 \nu^{*}_{jj} 
d_j^2 +  2 \nu^{*}_{12} d_1 d_2 \cos \frac{2 \pi \Delta}{\hbar \omega_c} \biggr],
\end{eqnarray} 
where the sums are taken over the subbands $j=1,2$ with energies $\varepsilon_j$ separated 
by $\Delta = |\varepsilon_2-\varepsilon_1|$. The second term is the first-order 
quantum correction describing the SdH oscillations ($\varepsilon_F= \hbar^2 \pi n_s/2m$ is 
the Fermi energy), and the third term is the equilibrium second-order 
quantum correction containing the MIS oscillations. The fourth and the fifth terms are non-equilibrium 
second-order quantum corrections describing the influence of microwaves owing to inelastic and 
displacement mechanisms, respectively. In Eq. (1), $\rho_{0}= m/e^2 n_s \tau_{tr}$, $\tau_{tr}$ is 
the averaged transport time defined as $1/\tau_{tr}=(\nu^{tr}_{1}+\nu^{tr}_{2})/2$, 
$1/\tau^*=(\nu^*_{1}+\nu^*_{2})/2$, $d_j=\exp(-\pi \nu_j/\omega_c)$ are the Dingle factors, 
${\cal T}= X/\sinh X$ with $X=2 \pi^2 T/\hbar \omega_c$ is the thermal suppression factor, 
and $n_j$ are the partial densities in the subbands ($n_1+n_2=n_s$). The subband-dependent 
quantum relaxation rates $\nu_j$ and $\nu_{jj'}$, as well as the scattering rates 
$\nu^{tr}_j$, $\nu^*_j$, $\nu^{tr}_{jj'}$, and $\nu^*_{jj'}$ are defined according to
\begin{eqnarray}
\nu_j=\sum_{j'=1,2} \nu_{jj'},~~~\nu^{tr}_j= \sum_{j'=1,2} \frac{n_j+n_{j'}}{n_s} \nu^{tr}_{jj'}, \nonumber \\ 
\nu^{*}_j=\sum_{j'=1,2} \left(\frac{n_j+n_{j'}}{n_s} \right)^2 \nu^{*}_{jj'},
\end{eqnarray}
and
\begin{eqnarray}
\begin{array}{c} \nu_{jj'} \\
\nu^{tr}_{jj'} \\ 
\nu^{*}_{jj'} \end{array}
 \left\} = \int_0^{2 \pi}
\frac{d \theta}{2 \pi} \nu_{jj'}(\theta) 
\times \right\{ \begin{array}{c} 1 \\
F_{jj'}(\theta) \\ 
F^2_{jj'}(\theta)  \end{array} , \\
\nu_{jj'}(\theta)=\frac{m}{\hbar^3} w_{jj'} \left( \sqrt{ (k^2_{j} +
k^2_{j'}) F_{jj'}(\theta)} \right), \nonumber
\end{eqnarray}
where $w_{jj'}(q)$ are the Fourier transforms of the correlators of the scattering potential, 
$F_{jj'}(\theta)=1 - 2 k_j k_{j'} \cos\theta/(k_j^2 + k_{j'}^2)$, $\theta$ is the scattering angle, 
and $k_j=\sqrt{2 \pi n_j}$ is the Fermi wavenumber for subband $j$. Next, 
\begin{equation}
A_{\omega} \simeq \frac{ {\cal P}_{\omega} (2 \pi
\omega/\omega_c) \sin (2 \pi \omega/\omega_c)}{1+{\cal P}_{\omega}
\sin^2(\pi \omega/\omega_c)}
\end{equation}
and 
\begin{equation}
B_{\omega} \simeq \frac{\tau_{tr}}{\tau^*} P_{\omega} \left[ \frac{\pi \omega}{\omega_c} \sin \frac{2 \pi \omega}{\omega_c} +\sin^2 \frac{\pi \omega}{\omega_c} \right]
\end{equation}
are dimensionless oscillating functions describing MIROs. The denominator of $A_{\omega}$ 
accounts for the saturation effect at high enough microwave intensity. Finally, 
\begin{equation}
{\cal P}_{\omega}=\frac{\tau_{in}}{\tau_{tr}} P_{\omega},~~P_{\omega} = 
\left(\frac{e E_{\omega} }{\hbar \omega}\right)^{2} \overline{v_F^2}
\frac{\omega^{2}_{c}+\omega^{2}}{\left(\omega^{2}-\omega^{2}_{c}\right)^{2}}.
\end{equation}
The dimensionless factor $P_{\omega}$ is proportional to the absorbed microwave power. $E_{\omega}$ is the 
amplitude of electric field of the microwaves, $\overline{v_F^2}=(v_1^2+v_2^2)/2$ is the averaged 
Fermi velocity (the Fermi velocities in the subbands are defined as $v_j=\hbar k_j/m$), and 
$\tau_{in}$ is the inelastic relaxation time. This expression for $P_{\omega}$ assumes linear 
polarization of microwaves and is valid away from the cyclotron resonance.

The general expression is considerably simplified in the case relevant to our DQWs, when 
$\Delta/2$ is much smaller than the Fermi energy $\varepsilon_F$. In this 
case one may approximate $n_1 \simeq n_2 \simeq n_s/2$ and $\nu_{11} \simeq \nu_{22}$, $\nu^{tr}_{11} 
\simeq \nu^{tr}_{22}$, $\nu^*_{11} \simeq \nu^*_{22}$, which leads also to $\nu_1 \simeq \nu_2$, 
$d_1 \simeq d_2$, $\nu^{tr}_{1} \simeq \nu^{tr}_{2} \simeq 1/\tau_{tr}$, and $\nu^{*}_{1} \simeq 
\nu^{*}_{2} \simeq 1/\tau^*$. Moreover, in balanced DQWs and under condition that interlayer correlation 
of scattering potentials is not essential, one has \cite{12} $\nu^{tr}_{12} \simeq \nu^{tr}_{jj}$ and 
$\nu^*_{12} \simeq \nu^*_{jj}$. Applying these approximations to Eq. (1), we rewrite it in the form
\begin{eqnarray}
\frac{\rho_{d}}{\rho_{0}} \simeq 1 - 2 {\cal T} d \sum_{j=1,2}
\cos \frac{2 \pi (\varepsilon_F -\varepsilon_j)}{\hbar \omega_c} \nonumber \\ 
+ d^2 \left[1-A_{\omega}-B_{\omega} \right] \left(1+ \cos \frac{2 \pi \Delta}{\hbar \omega_c} \right).  
\end{eqnarray} 
The second-order quantum contribution (the last term in this expression) is reduced to the 
corresponding single-subband form \cite{6} if the MIS oscillation factor $1+ \cos (2 \pi 
\Delta/\hbar \omega_c)$ is replaced by 2. The amplitude of this contribution is determined 
by the single squared Dingle factor $d^2=\exp(-2 \pi /\omega_c \tau_q)$, where the quantum 
lifetime is defined as $1/\tau_q \equiv (\nu_1+\nu_2)/2$. The MIROs are given by the term
$-A_{\omega}-B_{\omega}$ representing a combined action of the inelastic and displacement 
mechanisms. Since the factor $2 \pi \omega/\omega_c$ is large compared to unity in the region 
of integer MIROs ($\omega > \omega_c$), the functions $A_{\omega}$ and $B_{\omega}$ have 
nearly the same frequency dependence (if far from the saturation regime) and differ only by 
magnitude and by different sensitivity to temperature. 

The consideration presented above 
neglects the contribution of the photovoltaic mechanism, which, according to theory, 
should give a different frequency dependence leading, in particular, to a different 
phase of MIROs. This contribution decreases with increasing $\omega$. According to our 
theoretical estimates, the photovoltaic mechanism contribution in our samples can be 
neglected in comparison to contributions of both inelastic and displacement mechanisms 
at the frequencies we use, while in samples with higher mobilities its relative 
contribution is even smaller. Taking also into account that the phase shift in MIROs 
specific for the photovoltaic mechanism has not been detected experimentally, the neglect 
of this mechanism is reasonably justified.
  
For the analysis of experiments, we have to take into account the dependence of the 
characteristic scattering times: quantum lifetime $\tau_{q}$ and inelastic relaxation 
time $\tau_{in}$ on the effective electron temperature $T_{e}$. According to theory \cite{6}, 
based on consideration of electron-electron scattering, $\tau_{in}$ scales as 
\begin{equation} 
\frac{\hbar}{\tau_{in}} \simeq \lambda_{in} \frac{T^{2}_{e}}{\varepsilon_F},
\end{equation}
where $\lambda_{in}$ 
is a numerical constant of order unity. To take into account Landau level broadening 
owing to electron-electron scattering, a similar contribution should be added to 
inverse quantum lifetime \cite{16}, so $1/\tau_q$ is replaced with $1/\tau_q + 1/\tau_q^{ee} \equiv 
1/\tau_q(T_e)$, where $\hbar/\tau_q^{ee} \simeq \lambda T^{2}_{e}/\varepsilon_F$; the numerical 
constants $\lambda_{in}$ and $\lambda$ are not, in general, equal to each other. As a result, 
the Dingle factor becomes temperature-dependent: $d \rightarrow d(T_e)=\exp[-\pi /\omega_c 
\tau_q(T_e)]$.

For weak microwave power (far from the saturation regime), Eq. (7) can be rewritten in the form
\begin{eqnarray}
\frac{\rho_{d}}{\rho_{0}} \simeq 1 +  \left. \frac{\rho_{d}}{\rho_{0}} \right|_{S} +  
d^2(T_e)  \left(1+ \cos \frac{2 \pi \Delta}{\hbar \omega_c} \right) 
\biggl\{ 1- P_{\omega} \nonumber \\
\times \left[ \left( (T_0/T_e)^2 +\beta \right) \frac{2 \pi \omega}{\omega_c} 
\sin \frac{2 \pi \omega}{\omega_c} + 2 \beta \sin^2\frac{\pi \omega}{\omega_c}  \right] \biggr\},
\end{eqnarray}
where the SdH oscillation term from Eq. (7) is denoted as $\left. \rho_{d}/\rho_{0} \right|_{S}$.
In this expression we have applied the dependence $\tau_{in} \propto T_e^{-2}$
and denoted $T_0$ as the temperature when $\tau_{in}=\tau_{tr}$. Next, 
$\beta=\tau_{tr}/2 \tau^*$. The contributions proportional to $\beta$ come from the 
displacement mechanism. Since the first term in the square brackets is considerably 
larger than the second one, it dominates the frequency dependence of MIROs. Therefore, 
the combined action of the inelastic and displacement mechanisms on the magnetoresistance 
can be approximately described by using the expression for inelastic mechanism contribution 
with an effective (enhanced owing to the displacement mechanism) relaxation time $\tau^*_{in}$:
\begin{equation} 
\tau^*_{in} \equiv \tau_{in} +  \frac{\tau^2_{tr}}{2 \tau^*} = \tau_{tr} \left[(T_0/T_e)^2 +\beta \right]. 
\end{equation}
The crossover between inelastic and displacement mechanisms should take place at a 
characteristic temperature $T_C = T_0/\sqrt{\beta}$. Below we present our experimental 
results and compare them with the theoretical predictions. 

\section{Results and conclusions}

While similar results have been obtained for various frequencies between 55 and 140~GHz, 
we focus our analysis on the frequencies 85~GHz (attenuation -7.5~dB) and 110~GHz 
(attenuation 0~dB). The electric fields for both frequencies, $E_{\omega} =2$ V/cm 
(85~GHz, -7.5~dB) and $E_{\omega}=1.5$ V/cm (110~GHz, 0~dB), and the corresponding 
($B$-dependent) electron temperatures $T_e$ were estimated by comparing the effect of 
heating-induced suppression of SdH oscillations with a similar effect in the known 
dc electric fields. At low temperatures, these quantities are in agreement with those 
obtained by fitting calculated amplitudes of the magnetoresistance oscillations to 
experimental data. 

The theoretical magnetoresistance is calculated as explained above. The temperature 
dependence of quantum lifetime entering the Dingle factor is determined from 
temperature dependence of the MIS oscillations in the absence of microwaves (see the 
details in Ref. 11). This dependence fits well to the theoretically predicted 
one, where the contribution of electron-electron scattering enters with $\lambda =3.5$ 
(see previous section). The low-temperature quantum lifetime $\tau_q$ caused by impurity 
scattering is 3.5 ps, which corresponds to the ratio $\tau_{tr}/\tau_q \simeq 15$. The 
low-temperature magnetoresistance in the presence of microwave irradiation is satisfactory 
described by the inelastic mechanism contribution with $\tau_{in}$ of Eq. (8), and a 
comparison of experimental and theoretical results allows us to determine $\lambda_{in} 
\simeq 0.94$ in this dependence. 

\begin{figure}[ht]
\includegraphics[width=9cm]{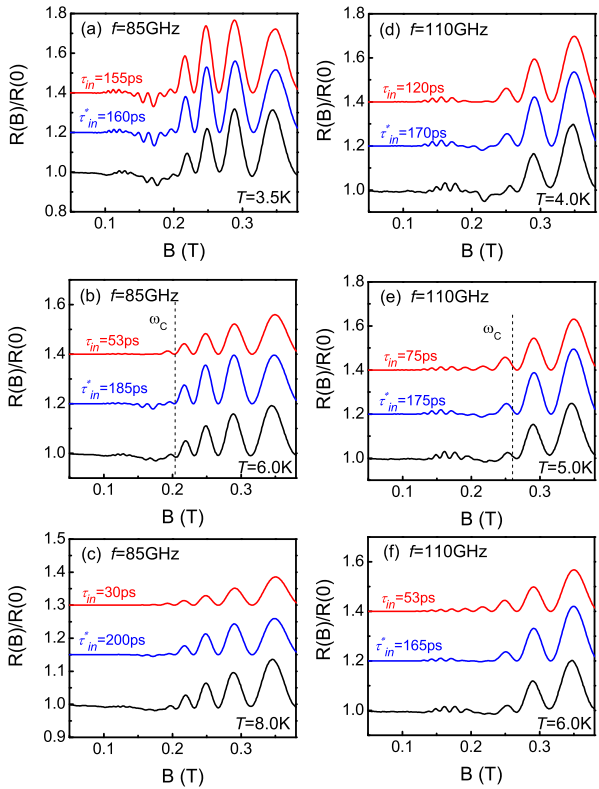}
\caption{\label{fig2}(Color online) Examples of measured and calculated magnetoresistance 
for 85~GHz (a-c) and 110~GHz (d-f). Red (top trace) is the theoretical magnetoresistance 
with the inelastic mechanism contribution. We display corresponding inelastic scattering 
time for the given electron temperature {$T=T_{e}$}. Blue (middle trace) is the theoretical 
magnetoresistance with an enhanced $\tau^{*}_{in}$, which fits the experimental data (black, 
bottom trace). Theoretical curves are shifted up for clarity.}
\end{figure} 

With increasing temperature, the inelastic mechanism alone fails to describe the experimental
magnetoresistance, and we have to introduce an enhanced relaxation time $\tau^{*}_{in}$. This 
is shown in Fig. \ref{fig2} where we plot the dc resistivity as a function of $B$ for the 
inelastic model with corresponding $\tau_{in}$ (red, top trace), inelastic model with an enhanced 
$\tau^{*}_{in}$ (blue, middle trace), and experimental trace (black, bottom trace) for several chosen temperatures. For 
both frequencies, the heating due to microwaves can be neglected for $T~\geq$~2.8~K, thus 
$T~\simeq~T_{e}$. It is clearly seen that with increasing temperature the theoretical model 
does not fit the magnetoresistance for 0.1~T $<$ B $<$ 0.3~T. Starting at 85~GHz [Fig. \ref{fig2}(a-c)] 
we find that neither the flipped MIS peaks around $B$~=~0.17~T nor the slightly enhanced MIS 
peaks at $B$~=~0.13~T occur in the inelastic model if we use calculated inelastic relaxation
time $\tau_{in}$. With an enhanced time $\tau^{*}_{in}$, e.g., in Fig. 
\ref{fig2}(b), with $\tau^{*}_{in}$~=3.5~$\tau_{in}$, both features appear at the corresponding 
magnetic field. This deviation is especially clear in Fig \ref{fig2}(c) at $T$~=~8~K. Here we 
use $\tau^{*}_{in}$~=6.7~$\tau_{in}$ to obtain the closest fit to the experimental result. For 
110~GHz, we observe similar results for all temperatures, and we show the features at $T$~=~4~K, 
5~K and 6~K, see Figs. \ref{fig2}(d-f). Due to a different frequency which changes strongly the MIS 
oscillation picture \cite{10}, we focus now on the enhanced MIS peaks around $B$~=~0.16~T 
and the damped features at $B$~=~0.22~T. Whereas the comparison with theoretical model only shows
a slightly smaller amplitude of the enhanced MIS peaks at $B$~=~0.16~T, the damped or inverted
MIS peaks [Fig. \ref{fig2}(d)] observed in experiment at $B$~=~0.22~T do not occur unless 
$\tau_{in}$ is enhanced to $\tau^{*}_{in}$. 

\begin{figure}[ht]
\includegraphics[width=9cm]{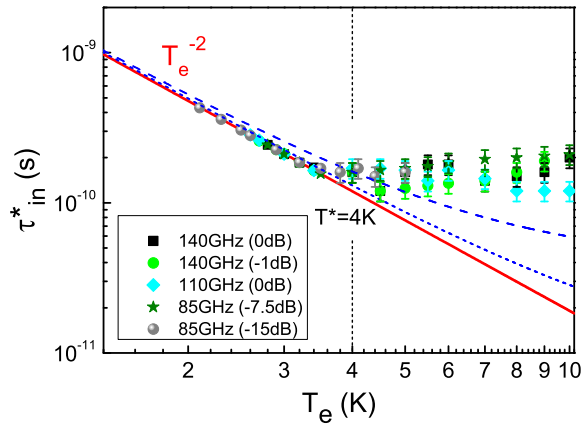}
\caption{\label{fig3}(Color online) Temperature dependence of the effective relaxation 
time $\tau^*_{in}$ extracted for different microwave frequencies and intensities (points), 
and theoretically predicted inelastic relaxation time $\tau_{in} \propto T_e^{-2}$ of Eq. (8) 
(red thick line). The deviation from the inelastic model starts at a critical temperature 
$T^{*}\simeq$~4~K. For higher $T_e$ we observe an almost temperature-independent behavior 
until the effect of microwaves on the DQW systems vanishes depending on the strength of 
the electric field $E_{\omega}$. The theoretical dependence of $\tau^*_{in}$ under 
approximations of smooth scattering potential (short dash) and of mixed disorder at 
maximal possible content of short-range scatterers (dash) are also shown.}
\end{figure} 	

In Fig. \ref{fig3} we show the enhanced relaxation time $\tau^{*}_{in}$ as a function 
of electron temperature $T_{e}$. We have added the data for a higher frequency 
of 140 GHz and for a lower microwave intensity (85~GHz at -15 dB, the estimated electric field 
is $E_{\omega}$~=~0.8~V/cm). It is clearly seen that $\tau^{*}_{in}$ is very close 
to $\tau_{in} \propto T_e^{-2}$ for $T_{e}~\leq~T^{*}$, which strongly confirms the 
relevance of the inelastic mechanism of photoresistance in this region of temperatures.
The deviation from this mechanism begins at $T^{*}\simeq$~4~K, which is identified as 
a ``critical" temperature. For $T_e > T^{*}$, a nearly temperature-independent (constant) 
$\tau^{*}_{in}$ is obtained in the whole frequency range. The dispersion of the 
experimental points in this region of temperatures is attributed to a limited accuracy 
of our analysis, when temperature dependence of the prefactor is extracted using the 
expressions containing temperature-dependent exponential factor $d^2(T_e)$. 
For each extracted $\tau^{*}_{in}$, we present an error bar in Fig. \ref{fig3} 
for $T>$3.5~K. Note that for low temperature the errors become smaller due to 
the $T^{-2}$-dependence of inelastic relaxation time.

It is tempting to attribute the observed behavior to the theoretically predicted 
crossover between the inelastic and displacement mechanisms. To check out the reliability
of this assumption, let us compare the experimental critical temperature $T^{*}$ with 
the theoretical crossover temperature. Based on our experimental data, we find 
$T_0 \simeq 6.0$ K. To find the parameter $\beta$, an additional consideration is required, 
since the time $\tau^*$ is not directly determined from experiment. This time is 
expressed through the angular harmonics of the scattering rate as \cite{17,18}  
\begin{equation} 
\frac{1}{\tau^*}=\frac{3}{2\tau_0}-\frac{2}{\tau_1}+\frac{1}{2\tau_2},
\end{equation}
while $1/\tau_q=1/\tau_0$ and $1/\tau_{tr}=1/\tau_0-1/\tau_1$. A large ratio of 
$\tau_{tr}/\tau_q$, which is typical for modulation-doped structures, suggests 
that the scattering is caused mostly by the long-range random potential (smooth 
disorder). If a model of exponential correlation is used [$w(q) \propto \exp(-l_c q)$, 
where $l_c$ is the correlation length of the random potential], each harmonic is given 
by the following expression:
\begin{equation} 
\frac{1}{\tau_k}=\frac{1}{\tau_{sm}} \frac{1}{1+ \chi k^2}, ~~~\chi=(k_F l_c)^{-2} \ll 1.
\end{equation}
Since the parameter $\chi$ can be determined from the known ratio $\tau_{tr}/\tau_q$, which 
is equal to $1+\chi^{-1}$ in this model, the time $\tau^*$ and, hence, $\beta$ can be found. 
For our samples we obtain the crossover temperature $T_C \simeq 15.3$ K, which is considerably 
larger than $T^{*}$. Therefore, the displacement mechanism contribution is not strong enough 
to explain the observed temperature behavior.

\begin{figure}[ht]
\includegraphics[width=9cm]{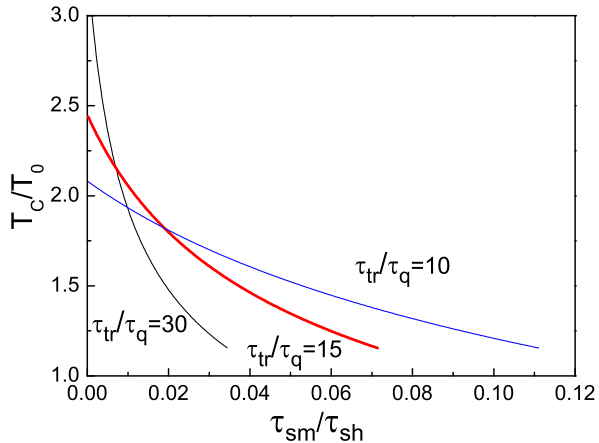}
\caption{\label{fig4}(Color online) Theoretical dependence of the crossover temperature 
on the content of short-range scatterers for several given ratios $\tau_{tr}/\tau_q$ (for 
our sample this ratio is 15). $T_0$ is the temperature when $\tau_{in}$ equals $\tau_{tr}$.}
\end{figure} 	

Recently, it was shown \cite{18} that the presence of a small amount of short-range 
scatterers (such as point defects whose radius is much smaller than the inverse 
Fermi wavenumber $1/k_F$) increases the contribution of the displacement mechanism. 
For this two-component disorder model, Eq. (12) should be replaced with \cite{17,18} 
\begin{equation} 
\frac{1}{\tau_k}=\frac{\delta_{k,0}}{\tau_{sh}} + \frac{1}{\tau_{sm}} \frac{1}{1+ \chi k^2}.
\end{equation}  
The relative content of the short-range scatterers can be characterized by the 
ratio $\tau_{sm}/\tau_{sh}$. The crossover temperature $T_C$, indeed, decreases 
with increasing $\tau_{sm}/\tau_{sh}$. However, to keep a constant $\tau_{tr}/\tau_q$ 
determined experimentally, one cannot make $\tau_{sm}/\tau_{sh}$ too large. 
In Fig. \ref{fig4}, we illustrate the dependence of $T_C/T_0$ on the content of the short-range 
scatterers for several ratios of $\tau_{tr}/\tau_q$. Each curve stops at the point when 
the given ratio cannot be reached if we add more short-range scatterers; this point 
corresponds to $\beta=3/4$. Therefore, for two-component disorder we can reduce $T_C$ 
down to $(2/\sqrt{3})T_0$, which in our case gives the lower limit $T_C \simeq 7$ K. Again, 
the displacement mechanism contribution is still weak to produce the crossover at 
$T \simeq 4$ K.  
 
To demonstrate the temperature dependence of the expected $\tau^*_{in}$, we add the 
theoretical plots based on Eq. (10) with $T_C=15.3$ K and $T_C = 7$ K to Fig. \ref{fig3}.  
It is clear that the smooth disorder model cannot fit the experimental data 
above $T^*=4$ K. The mixed disorder model produces a better (still not sufficient) 
agreement with experiment in this region, but leads to a noticeable deviation from the 
$\tau^*_{in} \propto T^{-2}$ dependence in the region $T < T^*$. This essential 
observation shows that the behavior of $\tau^*_{in}$ can hardly be explained within a 
model that adds a temperature-independent [as in Eq. (10)] or weakly temperature-dependent 
term to $\tau_{in}$: such a term cannot lead to a distinct change in the slope of the 
$T$-dependence around $T^*$. Therefore, one may suggest that another, previously 
unaccounted mechanism of photoconductivity, which turns on at $T \simeq T^*$ 
more abruptly than the displacement mechanism, should be important. 

In conclusion, we have studied the temperature dependence of magnetoresistance oscillations 
in the systems with two closely spaced 2D subbands (DQWs) under continuous microwave irradiation. 
With increasing temperature to $T^* \simeq 4$~K, we observe a considerable deviation from the 
temperature dependence predicted by the inelastic mechanism of microwave photoresistance. A similar 
behavior (at $T^{*} \simeq $2~K) has been recently observed in high-mobility quantum wells with one occupied 
subband \cite{9} and attributed to a crossover between inelastic and displacement mechanisms \cite{9,18}. 
We have analyzed our data in terms of this model, by taking into account elastic scattering of 
electrons by both long-range and short-range impurity potentials. We have found that even in the light 
of limited accuracy of our analysis, the observed deviation cannot be fully explained by the
contribution of the displacement mechanism, and, therefore, requires another explanation. We believe 
that this finding will stimulate further theoretical and experimental work on the transport 
properties of 2D electron systems exposed to microwave irradiation.   
 	
We thank M.A. Zudov and I.A. Dmitriev for useful discussions. This work was supported by 
COFECUB-USP (Project No. U$_{c}$~109/08), CNPq, FAPESP, and with microwave facilities 
from ANR MICONANO.

\end{document}